\documentstyle[pra,aps,twocolumn,psfig,floats,amssymb]{revtex}

\def\be{ \begin{equation} }
\def\ee{ \end{equation} }
\def\ba{ \begin{array} }
\def\ea{ \end{array} }
\def\bea{ \begin{eqnarray} }
\def\eea{ \end{eqnarray} }
\def\bml{ \begin{mathletters} }
\def\eml{ \end{mathletters} }
\def\bmla{ \bml \bea }
\def\emla{ \eea \eml }

\def\H{{\sf H}}
\def\U{{\sf U}}
\def\Ube{\U^{be}}
\def\Ubed{\U^{bed}}
\def\bright{ \psi_b }
\def\dark{ \psi_d }
\def\excited{ \psi_e}
\def\mixangle{ \alpha}
\def\phase{ \varphi}
\def\polar{ \vartheta}
\def\phaseL{ \beta}
\def\OmRMS{ \Omega}
\def\Cbe#1{\left[ \ba{c} c_b(#1) \\ c_e(#1) \ea\right]}
\def\P{{\cal P}}
\def\I{{\cal I}}
\def\Io{\I_0}
\def\N{{\cal N}}
\def\J{\N\P}
\def\Rcoh{\rho^{\text{coh}}}
\def\Rinc{\rho^{\text{inc}}}
\def\Ri{r^{\text{inc}}}
\def\Tr{{\text{Tr}}}
\def\Re{{\text{Re}}}
\def\Im{{\text{Im}}}
\def\Pe{P_e}

\begin{document}

\draft

\wideabs{

\title{Measuring a coherent superposition}

\author{N. V. Vitanov$^1$\thanks{e-mail: vitanov@rock.helsinki.fi},
B. W. Shore$^{2,3}$, R. G. Unanyan$^{2,4}$, and K. Bergmann$^2$}

\address{
$^1$Helsinki Institute of Physics, PL 9,
 00014 Helsingin yliopisto, Finland\\
$^2$Fachbereich Physik der Universität, 67653 Kaiserslautern, Germany \\
$^3$Permanent address: Lawrence Livermore National Laboratory,
 Livermore, CA 94550, USA \\
$^4$Permanent address: Institute for Physical Research,
 Armenian National Academy of Sciences, 378410 Ashtarak-2, Armenia
}

\date{\today }

\maketitle

\begin{abstract}
We propose a simple method for measuring the populations and the relative
phase in a coherent superposition of two atomic states.
The method is based on coupling the two states to a third common (excited)
state by means of two laser pulses, and measuring the total fluorescence
from the third state for several choices of the excitation pulses.
\end{abstract}


}


\section{Introduction}

Atoms and molecules prepared in well-defined coherent superpositions
of energy states exhibit many interesting properties, such as
dark resonances
\cite{Arimondo96},
subrecoil laser cooling
\cite{Aspect88,Lawall94LC,Lawall95,Cohen-Tannoudji90},
electromagnetically induced transparency
\cite{Hakuta91,Boller91,Field91},
radiation amplification without population inversion
\cite{Kocharovskaya92,Scully92,Nottelmann93,Veer93},
refractive index enhancement without absorption
\cite{Scully91},
and enhanced harmonic generation
\cite{Gauthey95,Watson96}.
Coherent superpositions are essential to the implementation of quantum
computation and quantum cryptography or, more generally,
quantum information
\cite{Williams97,Steane98}.

Various techniques are available for preparing coherent superpositions.
Some of them are sensitive to pulse fluence (the time integrated pulse
area), e.g.,
radio frequency  or microwave excitation
\cite{Fill90,Kocharovskaya92a,Bitto93},
resonant optical pulses
\cite{Veer93},
and trains of identical pulses
\cite{Nottelmann93}.
Techniques based on stimulated Raman adiabatic passage (STIRAP)
\cite{Kuklinski89,Gaubatz90,Bergmann98}
are  relatively insensitive to pulse area, e.g., fractional STIRAP
\cite{Marte91,Lawall94BS,Weitz94PRA,Weitz94PRL,Vitanov99}
and tripod-linkage adiabatic passage
\cite{Unanyan98}.

To verify the reliability of these techniques
it is essential to have a method for measuring the parameters
of the created coherent superposition --
the populations, the relative phase between the two states,
and the degree of coherence.
In this paper, we propose such a method.
It is based on coupling the two states comprising the superposition
to a third state (excited and subject to radiative decay), by means
of two laser pulses, thus mapping the superposition parameters onto
the population of the excited state.
This population could then be observed either by fluorescence,
by photoionization, or by recording with a channeltron the presence
of an excited atom impinging upon a sensitive surface.
By measuring the total signal from the excited state (for simplicity
we shall refer to it as fluorescence) for four different combinations
of laser polarizations one can deduce unambiguously the superposition
parameters.

The method applies to superpositions of states that are unconnected by
direct dipole coupling, but are linked by a two-photon transition.
An example occurs with the frequently used superposition between
the $m=-1$ and $m=+1$ magnetic sublevels of a degenerate level
having angular momentum $J=1$.
In the cases of fractional STIRAP
\cite{Marte91,Lawall94BS,Weitz94PRA,Weitz94PRL,Vitanov99}
and in the tripod scheme
\cite{Unanyan98},
the excited state can be the one used in creating the superposition;
no additional lasers are then required to implement the proposed
superposition measurement.

This paper is organized as follows.
In Sec.~\ref{Sec-pure} we consider the ideal case when the initial
superposition is a {\em pure} state, i.e., there is no incoherent
population in the superposition.
This allows us to present some of the fundamental principles
in a simple way.
In Sec.~\ref{Sec-mixed} we consider the general case when the
initial superposition is a {\em mixed} state.
This case is physically more realistic because there is usually
a nonzero probability that during the preparation of the coherent
superposition some of the atoms are pumped incoherently
(e.g., by spontaneous emission) into one of the two states involved
in the superposition.
Finally, in Sec.~\ref{Sec-conclusion} we summarize the conclusions.


\section{Measuring a {\em completely} coherent superposition (pure state)}

\label{Sec-pure}

\subsection{Theoretical background}

\subsubsection{Superposition parameters: Bloch and Poincar\`e spheres}


We assume that the atom is prepared initially in a coherent
superposition of two states $\psi_1$ and $\psi_2$,
\be \label{superposition}
\Psi = \psi_1 \cos \mixangle + \psi_2 e^{i\phase} \sin \mixangle,
\ee
and we wish to determine the unknown superposition parameters --
the mixing angle $\mixangle$ ($0\leq \mixangle \leq \pi /2$)
and the relative phase $\phase$ between $\psi_1$ and $\psi_2$
($-\pi < \phase \leq \pi $).

The parametrization of the superposition (\ref{superposition}) by two
angles is closely related to the parametrization of the Bloch vector,
whose components are bilinear combinations of probability amplitudes.
The dynamics of two-state excitation is often visualized by the motion
of this vector on the Bloch sphere.
Because the Bloch vector length is set to unity for coherent excitation,
the state of a coherent system is completely fixed by two angles.
The azimuthal and polar angles of the Bloch vector for state
(\ref{superposition}) are given by $\pi-2\mixangle$ and $-\phase$,
respectively.


This parametrization of a two-state superposition is similar to
the parametrization of polarized light by means of two angles that
locate a point on the Poincar\`e sphere \cite{Born70,Klein70,Shore90}.
In the latter case there exist an infinite number of choices
for the two independent orthogonal polarizations needed
to characterize polarized light;
orthogonal pairs lie at opposite poles on the sphere.
In the case of coherent superpositions of nondegenerate states there
is a natural basis -- that of the two distinct energy states.
When the atomic states are degenerate, as happens with magnetic
sublevels, then the quantization axis -- and the consequent definition
of basis states -- is arbitrary.
The analogy between pairs of orthogonal polarization states and
orthogonal atomic states is then complete.


\subsubsection{Mapping the superposition parameters
onto the population of an excited state}

\begin{figure}[tb]
\vspace*{0mm}
\centerline{\psfig{width=32mm,file=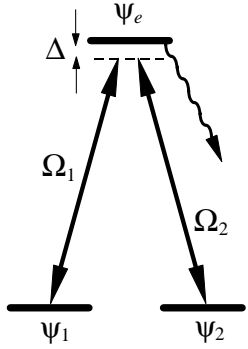}}
\vspace*{4mm}
\caption{
Sketch of the proposed scheme for measuring the coherent superposition
of states $\psi_1$ and $\psi_2$.
The superposition parameters (the populations and the relative phase)
are mapped onto the population of the excited state $\excited$ by
coupling state $\psi_1$ to $\excited$ with the laser pulse $\Omega_1$
and state $\psi_2$ to $\excited$ with the laser pulse $\Omega_2$.
The excitation pulses together obey a
two-photon Raman-like resonance condition, although the individual pulses
need not be resonant with their respective Bohr frequencies.
The single-photon detuning is $\Delta$.
}
\label{Fig-system}
\end{figure}

In order to determine the superposition parameters $\mixangle$ and
$\phase$, we propose to use a pair of laser pulses, each of which
couples one of the two superposed states $\psi_1$ and $\psi_2$ to
a third (excited) state $\excited$, as shown in Fig.~\ref{Fig-system}.
The excitation produced by these pulses {\em maps} the initial
superposition parameters onto the population of state $\excited$,
thereby reducing the measurement of a coherent superposition
to the easier measurement of population.
Moreover, we shall show below that, by measuring the total fluorescence
signal from the excited state for four different discrete combinations
of laser parameters, one can determine unambiguously the parameters of
the initial superposition (\ref{superposition}), without the necessity
of knowing the exact probability of transition to the excited state.

In the mathematical description of this scheme, we assume
applicability of the usual rotating-wave approximation (RWA),
and take the time-varying interaction between states $\psi_1$
and $\excited$ to be the Rabi frequency $\Omega_1(t)$,
while the coupling between states $\psi_2$ and $\excited$
is described by the Rabi frequency $\Omega_2(t)$.
We require the two Rabi frequencies to have the same (pulsed) time
dependence, as described by the envelope function $f(t/T)$
(with $T$ being the pulse width),
but we allow the fields to have different phases $\phaseL_1$ and
$\phaseL_2$ and different peak Rabi frequencies $A_1$ and $A_2$,
\be \label{pulses}
\Omega_1(t) = A_1 f(t/T), \qquad \Omega_2(t) = A_2 f(t/T).
\ee
We assume that the  carrier frequencies for the two pulses
together obey a
two-photon Raman-like resonance condition, but we allow
the possibility of nonzero single-photon detuning $\Delta$.
The RWA probability amplitudes of the three states obey
the Schrödinger equation \cite{Shore90},
\be \label{3SS}
i\hbar \frac{d{\bf c}(t)}{dt}={\H}(t){\bf c}(t),
\ee
where ${\bf c}(t)=\left[ c_1(t),c_e(t),c_2(t)\right] ^T$
and $\H(t)$ is the RWA Hamiltonian,
\be \label{H}
{\H}(t) = \frac{\hbar}2
\left[ \ba{ccc}
0 & \Omega_1(t)e^{-i\phaseL_1} & 0 \\
\Omega_1(t)e^{i\phaseL_1} & 2\Delta  & \Omega_2(t)e^{-i\phaseL_2} \\
0 & \Omega_2(t)e^{i\phaseL_2} & 0
\ea \right].
\ee
We wish to find the population $\Pe =\left|c_e(+\infty)\right|^2$ of
the excited state $\excited$ after the excitation, provided the system
has been initially in the superposition state (\ref{superposition}).

Due to the two-photon resonance and the identical time dependence of
$\Omega_1(t)$ and $\Omega_2(t)$, the three-state dynamics is reduced
to that of a two-state system \cite{Vitanov98}.
This is achieved by replacing states $\psi_1$ and $\psi_2$ with two
alternative states, a ``bright'' state $\bright$ and a ``dark'' state
$\dark$, which are linear combinations of $\psi_1$ and $\psi_2$,
\bmla
\label{bright}
&& \bright = \psi_1 e^{-i\phaseL_1} \sin\polar
	  + \psi_2 e^{ i\phaseL_2} \cos\polar, \\
\label{dark}
&& \dark = \psi_1 e^{-i\phaseL_2} \cos\polar
	  -\psi_2 e^{ i\phaseL_1} \sin\polar,
\emla
where the constant angle $\polar$ is defined by
\be
\label{theta}\tan \polar =\frac{A_1}{A_2},
	\qquad (0\leq \polar \leq \pi/2).
\ee
We stress that the bright and dark states are defined by the laser
parameters and not by the initial atomic state.

In the bright-excited-dark basis, the dark state is decoupled from
the other two states and its amplitude is conserved,
$c_d(t)=\text{const}$.
The three-state dynamics is thus reduced to a two-state one involving
the bright and excited states whose amplitudes obey the equation
\be \label{2SS}
i \frac d{dt} \Cbe{t} = \case12
 \left[ \ba{cc} 0 & \OmRMS(t) \\ \OmRMS(t) & 2\Delta \ea \right]
 \Cbe{t}.
\ee
It is important that the two-state Hamiltonian in Eq.~(\ref{2SS})
depends only on the rms Rabi frequency,
\be
\OmRMS(t)=\sqrt{\Omega_1(t)^2+\Omega_2(t)^2},
\ee
but not on the laser polarization $\polar$ and phases $\phaseL_1$ and
$\phaseL_2$.

The (unitary) transition matrix $\Ube$ for the effective two-state
problem (\ref{2SS}), which is defined by
\be
 \Cbe{+\infty} = \Ube \Cbe{-\infty},
\ee
can be parametrised by two complex numbers $a$ and $b$ as
\be \label{U2}
\Ube =
\left[ \ba{cc} a & b \\ -b^* & a^* \ea \right].
\ee
Then the transition matrix for the three-state problem in the
bright-excited-dark basis is
\be \label{Ubed}
\Ubed =
\left[ \ba{ccc} a & b & 0 \\ -b^* & a^* & 0 \\ 0 & 0 & 1 \ea \right].
\ee
By using the transformation back to the bare-state basis,
we find that the transition matrix for the original three-state
problem (\ref{3SS}) is given in terms of $a$ and $b$ as
\be \label{U}
\U =\! \left[ \!\ba{ccc}
a\sin^2\polar + \cos^2\polar & \!b e^{-i\phaseL_1}\sin\polar &
\frac12 (a\!-\!1)e^{-i\phaseL}\sin 2\polar \\
-b^*e^{i\phaseL_1}\sin\polar & a^* & -b^*e^{-i\phaseL_2}\cos\polar \\
\frac12(a\!-\!1)e^{i\phaseL}\sin2\polar & be^{i\phaseL_2}\cos\polar &
a\cos^2\polar + \sin^2\polar
\ea \! \right] \!\!,
\ee
with $\phaseL = \phaseL_1 + \phaseL_2$.
By applying $\U$ on the initial vector (\ref{superposition}), we find
that the final population of state $\excited$ is 
\be \label{Pe}
\Pe = p\left| \sin\polar \cos\mixangle \
 + e^{i(\phase-\phaseL)}
	\cos\polar \sin\mixangle \right| ^2,
\ee
where $p=\left| b\right| ^2$ is the transition probability
in the equivalent two-state problem.
For resonant excitation ($\Delta=0$) we have
$p = \sin^2 \int_{-\infty}^{+\infty}\OmRMS(t)dt$,
whereas when $\Delta \neq 0$ the probability $p$ depends
on the detuning, on the pulse areas, and on the pulse shape $f(t/T)$.
An important feature of our method is that we do {\it not} need the precise
probability $p$ because the dependence of $\Pe$ on $p$ is factorized
(this does not apply to $P_1$ and $P_2$!).
Hence, $p$ can be eliminated simply by measuring $\Pe$ for
different laser parameters and taking ratios.


\subsection{{\em Detecting} a coherent superposition}

The $\phase$-dependence of the excited-state population (\ref{Pe}) suggests
a straightforward way of proving that a certain superposition is coherent.
After the preparation of the superposition and prior to the application of
the measuring laser fields, one can alter the superposition phase $\phase$
(e.g., by using pulsed magnetic field or off-resonance interaction) and plot
the total fluorescence signal as a function of the parameter that alters
the phase, e.g., the magnetic field $B$.
A non-constant dependence of $\Pe$ on $B$ is the indication of coherence.


\subsection{{\em Measuring} a coherent superposition}

Measuring the values of the superposition parameters $\mixangle$ and $\phase$
is more complicated that just proving coherence.
The result (\ref{Pe}) suggests two possible methods of measuring
$\mixangle$ and $\phase$.


\subsubsection{Polarization measurement}

Equation (\ref{Pe}) shows that $\Pe$ vanishes when $\polar = \mixangle$ and
$\phaseL = \phase + \pi$.
In this case the superposition (\ref{superposition})
coincides with the dark state.
Hence, we can measure the superposition parameters $\mixangle$ and $\phase$
by adjusting the ratio of the laser field amplitudes and the phase $\phaseL$
until the fluorescence from state $\excited$ disappears.

In the case when states $\psi_1$ and $\psi_2$ are the $m=-1$ and $m=+1$
magnetic sublevels of a $J=1$ level and state $\excited$ is the $m=0$
sublevel of a
$J=0$ or $J=1$ level
\cite{Vitanov99,Unanyan98},
this can be done by using a single elliptically polarized laser pulse,
which can be seen as a superposition of two circularly polarized
$\sigma^+$ and $\sigma^-$ pulses.
The electric field of the elliptically polarized pulse in the complex
representation $E(t)=E_x(t)+iE_y(t)$ is given by
\cite{Born70,Klein70,Shore90,Suominen91}
\be \label{elliptic}
E(t) = E_1(t) e^{-i\omega t + i \phaseL_1}
     + E_2(t) e^{ i\omega t + i \phaseL_2}.
\ee
The first term represents the $\sigma^+$ component and the second term is
the $\sigma^-$ component; hence, $E_1(t)/E_2(t)=A_1/A_2=\tan\polar$.
Here $\frac12\phaseL=\frac12(\phaseL_1+\phaseL_2)$
 is the angle of rotation
of the polarization ellipse and $|E_1-E_2|/(E_1+E_2)$
is its axial ratio
\cite{Born70,Klein70,Shore90,Suominen91}.
Measuring the superposition parameters in this manner, however, represents
a two-dimensional optimization procedure which may be inconvenient,
inefficient or inaccurate.
We propose below an alternative method.


\subsubsection{Population measurement}

The alternative method consists of measuring several total fluorescence
signals $\Io(\polar,\phaseL)$ (the subscript zero stands for the fact
that the initial superposition does not contain any incoherence) from
state $\excited$ for various sets of pulse pairs with different laser
polarizations $\polar$ and phases $\phaseL$.
Because fluorescence is proportional to the excited-state population,
$\Io(\polar,\phaseL) = \N \Pe(\polar,\phaseL)$ ($\N$ being essentially
the number of atoms), Eq.~(\ref{Pe}) suggests that it is convenient to
make the following set of measurements:
\bml
\bea
\label{Pe-1}
\Io(0,0) &=& \N p\sin^2\mixangle,\\
\label{Pe-2}
\Io(\pi/2,0) &=& \N p\cos^2\mixangle,\\
\label{Pe-3}
\Io(\pi/4,0) &=& \case12 \N p [1 + \sin 2\mixangle \cos\phase],\\
\label{Pe-4}
\Io(\pi/4, \pi/2) &=& \case12 \N p [1 + \sin 2\mixangle \sin\phase].
\eea
\eml
In the case when states $\psi_1$ and $\psi_2$ are the $m=-1$ and $m=+1$
sublevels of a $J=1$ level and one uses a single elliptically polarized
laser pulse with the field (\ref{elliptic}),
the first two cases (\ref{Pe-1}) and (\ref{Pe-2}) correspond to
$\sigma^+$ or $\sigma^-$ polarizations, respectively.
The third and fourth cases correspond to linear polarizations,
the latter being rotated to 45 degree with respect to the former.

The parameters of the initial superposition (\ref{superposition}) can
easily be derived from the fluorescence signals as follows:
\bml
\label{parameters0}
\bea
\tan^2\mixangle &=& \frac{\Io(0,0)}{\Io(\pi/2,0)},\\
\cos\phase &=& \frac{2\Io(\pi/4,0)-\Io(0,0)-\Io(\pi/2,0)}
		    {2\sqrt{\Io(0,0)\Io(\pi/2,0)}},\\
\sin\phase &=& \frac{2\Io(\pi/4,\pi/2)-\Io(0,0)-\Io(\pi/2,0)}
		    {2\sqrt{\Io(0,0)\Io(\pi/2,0)}}.
\eea
\eml
Knowing $\tan^2\mixangle$ is sufficient for the determination of
$\mixangle$ because $0\leq\mixangle\leq\pi/2$.
In order to find $\phase$, however, we need to know both $\cos\phase$
and $\sin\phase$ because $\phase$ is defined in the interval
$(-\pi,\pi]$.
Hence, we need {\em four} separate measurements of the total
fluorescence from the excited state $\excited$.
This method appears easier to implement than the two-dimensional
optimization discussed above.

The set of four measurements taken with different settings of two
angles is reminiscent of the settings of phase retardation angles
in the measurement of Stokes parameters \cite{Born70};
the connection will be further discussed in Sec.~\ref{Sec-mixed}.

Finally, as Eq.~(\ref{Pe}) shows, $\Pe$ depends on the detuning
and on the laser intensity through $p$ only.
Hence, the superposition parameters $\mixangle$ and $\phase$ obtained
by either methods described above should not depend on laser power,
detuning and pulse shape.


\section{Measuring a {\em partially} coherent superposition
(mixed state)}

\label{Sec-mixed}

\subsection{Density matrix description: analogy with partially
 polarized light}

The most general presentation of the properties of a two-state system
is by means of its density matrix.
The diagonal elements are real (and non-negative) and sum to unity,
and the off-diagonal elements are complex conjugates of each other,
so that a total of three real numbers suffice to completely specify
a two-state system.
The additional parameter, beyond the two parameters needed to describe
a coherent superposition, expresses the degree of incoherence between
the two states.

The density matrix description of a two-state atom has an analog
in the use of a coherence matrix to describe partially polarized light
as a mixture of elliptically polarized light
and unpolarized light.
A common parametrization of such light is by means of the four Stokes
parameters $s_j$ $(j=0,1,2,3)$, which can be regarded as the
coefficients in the expansion of the density matrix
in terms of the Pauli matrices and the unit matrix.
One of the Stokes parameters ($s_0$) measures the trace of the coherence
matrix (or overall intensity), and is not of interest in the present
context.

As is discussed in standard textbooks, the determination of the four
Stokes parameters can be accomplished with a set of six intensity
measurements, each involving a projection onto a pure polarization:
a pair of $\sigma^+$ and $\sigma^-$ circular polarizations, a pair of
$x$ and $y$ linear polarizations, and a pair of linear polarizations
rotated at 45 degree.
Three of the Stokes parameters can be regarded as associated with
projection of the polarization onto the three choices of (complex)
unit vectors: having introduced such a projection onto a particular
pair of orthogonal polarizations, the Stokes parameter is
the difference between the two orthogonal components.

A similar procedure can be applied to the case of two degenerate atomic
states linked, by electric dipole transitions, to an excited state.
The basic idea is to project the two-dimensional space of atomic states
onto various orthogonal axes, defined by the polarization of the
electric field that occurs in the interaction $-{\bf d \cdot E}$.
The measurement pulses map these projections onto the population
of the excited state $\excited$.


It should be emphasized that two of the six polarization measurements
are redundant.
Indeed, in the method we describe below only four independent
measurements of the fluorescence from the excited state are required.





\subsection{Mapping the superposition parameters
onto the excited-state population}

The time evolution of the density matrix $\rho \equiv \rho(t)$
 of the three-state system obeys the Liouville equation
\be\label{RhoEq}
i\hbar \dot{\rho}=\left[ \H, \rho \right],
\ee
where $\H(t)$ is the RWA Hamiltonian (\ref{H}) and
\be\label{Rho}
\rho = \left[ \ba{ccc}
\rho_{11} & \rho_{1e} & \rho_{12} \\
\rho_{e1} & \rho_{ee} & \rho_{e2} \\
\rho_{21} & \rho_{2e} & \rho_{22}
\ea \right].
\ee
Explicitly, the density-matrix equations read
\bml\label{Eqs-12e}
\bea
i\dot{\rho}_{11} &=&\Omega_1 \left( e^{-i\phaseL_1}\rho_{e1}-e^{i\phaseL_1}
 \rho_{1e}\right), \\
i\dot{\rho}_{22} &=&\Omega_2
 \left( e^{i\phaseL_2}\rho_{e2}-e^{-i\phaseL_2}\rho_{2e}\right), \\
i\dot{\rho}_{ee} &=&\Omega_1
 \left( e^{i\phaseL_1}\rho_{1e}-e^{-i\phaseL_1}\rho_{e1}\right) \nonumber\\
&+&\Omega_2\left( e^{-i\phaseL_2}\rho_{2e}-e^{i\phaseL_2}\rho_{e2}\right),\\
i\dot{\rho}_{12} &=&\Omega_1 e^{-i\phaseL_1}\rho_{e2}
 -\Omega_2 e^{-i\phaseL_2}\rho_{1e}, \\
i\dot{\rho}_{1e} &=&\Omega_1
 e^{-i\phaseL_1}\left( \rho_{ee}-\rho_{11}\right)
 -\Omega_2 e^{i\phaseL_2}\rho_{12}-\Delta \rho_{1e}, \\
i\dot{\rho}_{2e} &=&\Omega_2 e^{-i\phaseL_2}
 \left( \rho_{ee}-\rho_{22}\right) -\Omega_1
  e^{-i\phaseL_1}\rho_{21}-\Delta \rho_{2e}.
\eea
\eml

As in the fully coherent case (Sec.~\ref{Sec-pure}), the equations of
motion simplify considerably in the bright-excited-dark basis;
there the density-matrix elements read
\bml\label{Rho-bde}
\bea
\rho_{bb} &=& \rho_{11}\sin^2\polar +\rho_{22}\cos^2\polar \nonumber\\
&& +\left( e^{i\phaseL }\rho_{12}
 + e^{-i\phaseL }\rho_{21}\right) \sin \polar \cos \polar, \\
\rho_{dd} &=& \rho_{11}\cos^2\polar +\rho_{22}\sin^2\polar \nonumber\\
&& -\left( e^{i\phaseL }\rho_{12}
 + e^{-i\phaseL} \rho_{21}\right) \sin \polar \cos \polar, \\
\rho_{bd} &=& (\rho_{11}-\rho_{22}) \sin \polar \cos \polar
 e^{i(\phaseL_1-\phaseL_2)} \nonumber\\
&& -\rho_{12} e^{ 2i\phaseL_1} \sin^2 \polar
   +\rho_{21} e^{-2i\phaseL_2} \cos^2 \polar, \\
\rho_{be} &=& \rho_{1e} e^{i\phaseL_1} \sin \polar
 +\rho_{2e} e^{-i\phaseL_2} \cos \polar, \\
\rho_{de} &=& \rho_{1e} e^{i\phaseL_2} \cos \polar
 - \rho_{2e} e^{-i\phaseL_1} \sin \polar,
\eea
\eml
The six coupled density-matrix equations (\ref{Eqs-12e}) decompose into
three uncoupled sets of equations:
one equation expressing the conservation of the dark-state population,
\be\label{Eq-d}
i\dot{\rho}_{dd}=0;
\ee
two coupled equations for the dark-state coherences,
\bml \label{Eqs-dbe}
\bea
i\dot{\rho}_{db} &=&\OmRMS \rho_{de},  \\
i\dot{\rho}_{de} &=&-\OmRMS \rho_{db}-\Delta \rho_{de};
\eea
\eml
and, as befits the two-state dynamics, three coupled equations
involving the bright and excited states,
\bml\label{Eqs-be}
\bea
i\dot{\rho}_{bb} &=&\OmRMS\left( \rho_{eb}-\rho_{be}\right), \\
i\dot{\rho}_{ee} &=&\OmRMS\left( \rho_{be}-\rho_{eb}\right), \\
i\dot{\rho}_{be} &=&\OmRMS\left( \rho_{ee}-\rho_{bb}\right)
 -\Delta \rho_{be}.
\eea
\eml

Because the fluorescence signal is proportional to the excited-state
population $\rho_{ee}$, we are only interested in the third set of
equations (\ref{Eqs-be}).
It is important that Eqs.~(\ref{Eqs-be}) depend only on $\OmRMS$ and
$\Delta$, but not on the laser polarization $\polar$ and phases
$\phaseL_1$ and $\phaseL_2$.
The {\em initial conditions}, though, depend on $\polar$ and
$\phaseL=\phaseL_1+\phaseL_2$; they are
\bml\label{Rho-be}
\bea
\rho_{bb}(-\infty) &=&\rho_{11}(-\infty)\sin^2\polar
 +\rho_{22}(-\infty)\cos^2\polar\nonumber\\
&& \!\!\!\!\!\!\!\!\!\!\!\!\!\!\!\!\!
 +\left[e^{i\phaseL}\rho_{12}(-\infty)+e^{-i\phaseL }\rho_{21}(-\infty)\right]
 \sin \polar \cos \polar ,
\label{Rho-bb} \\
\rho_{ee}(-\infty) &=& \rho_{be}(-\infty) = 0.
\eea
\eml
(as the system is initially in a superposition of states
$\psi_1$ and $\psi_2$ only, all density-matrix elements involving
$\excited$ vanish initially).
It follows from Eqs.~(\ref{Rho-be}) and from the fact that
Eqs.~(\ref{Eqs-be}) are linear differential equations that
the excited-state population $\rho_{ee}$ is expressible as
\be\label{fluorescence}
\rho_{ee}(+\infty) = \P \rho_{bb}(-\infty),
\ee
where the probability $\P$ depends only on $\OmRMS(t)$ and $\Delta$,
but not on $\polar$ and $\phaseL$.
Hence, when we vary the polarization and the phases of the two laser
fields, while keeping $\OmRMS(t)$ and $\Delta$ unchanged,
the change in the fluorescence signal will derive entirely from
the change in the initial bright-state population $\rho_{bb}(-\infty)$.
This factorization (\ref{fluorescence}) of $\rho_{ee}(+\infty)$ is very
important because it allows us to avoid the necessity of knowing the
probability $\P$ and thus eliminates the dependence on $\OmRMS(t)$
and $\Delta$.


\subsection{Determination of the density matrix}

Equations (\ref{Rho-bb}) and (\ref{fluorescence}) suggest a simple
procedure for the determination of the initial density-matrix elements
by measuring the total fluorescence $\I(\polar,\phaseL)$
from the excited state.
We need only {\em four measurements} for different polarization
parameters $\polar$ and $\phaseL$.
As in Sec.~\ref{Sec-pure}, a suitable choice is the following set
of measurements:
\bml\bea
\I(0,0) &=& \J \rho_{22}(-\infty), \\
\I(\pi /2,0) &=& \J \rho_{11}(-\infty), \\
\I(\pi /4,0) &=& \case12 \J [ \rho_{11}(-\infty) + \rho_{22}(-\infty)
 \nonumber\\
&+& 2\Re \rho_{12}(-\infty)], \\
\I(\pi /4,\pi /2) &=& \case12 \J [ \rho_{11}(-\infty) + \rho_{22}(-\infty)
 \nonumber\\
&-& 2\Im \rho_{12}(-\infty)],
\eea\eml
where we have used the relation
$\I(\polar,\phaseL)=\N\Pe(\polar,\phaseL)$.
Because $\rho_{11}(-\infty) + \rho_{22}(-\infty) = 1$,
we have $\J = \I(0,0) + \I(\pi /2,0)$.
Hence the initial density-matrix elements are given by
\bml
\bea
\rho_{11}(-\infty) &=& \frac{\I(0,0)}{\I(0,0)+\I(\pi /2,0)}, \\
\rho_{22}(-\infty) &=& \frac{\I(\pi /2,0)}{\I(0,0)+\I(\pi /2,0)}, \\
\Re \rho_{12}(-\infty) &=& \frac{\I(\pi/4,0)}{\I(0,0)+\I(\pi /2,0)}
	- \case12, \\
\Im \rho_{12}(-\infty) &=& \case12
	- \frac{\I(\pi/4,\pi/2)}{\I(0,0)+\I(\pi /2,0)} .
\eea
\eml


\subsection{Determination of the superposition parameters}

Once we know the density matrix, it is easy to find its incoherent part
$\Rinc$ and its coherent part $\Rcoh$ by using the decomposition
(we drop the argument $-\infty$ hereafter)
\be
\rho = \Rinc + \Rcoh
     = \left[\ba{cc} \Ri & 0 \\ 0 & \Ri \ea\right]
	+ \left[\ba{cc} \Rcoh_{11} & \Rcoh_{12} \\ 
		        \Rcoh_{21} & \Rcoh_{22} \ea\right],
\ee
and the coherence relation
$\Rcoh_{11} \Rcoh_{22} - \Rcoh_{12} \Rcoh_{21} = 0$.
The explicit expressions are readily found \cite{Born70},
\bml
\bea
\Rcoh_{11} &=& \case12 \left[ \rho_{11} - \rho_{22}
 + \sqrt{(\rho_{11}-\rho_{22})^2 + 4\left|\rho_{12}\right|^2} \right],\\
\Rcoh_{22} &=& \case12 \left[ \rho_{22} - \rho_{11}
 + \sqrt{(\rho_{11}-\rho_{22})^2 + 4\left|\rho_{12}\right|^2} \right],\\
\Rcoh_{12} &=& \rho_{12}, 
\eea
\eml
\be
 \Ri = \case12 \left[ \rho_{11} + \rho_{22}
 - \sqrt{(\rho_{11}-\rho_{22})^2 + 4\left|\rho_{12}\right|^2} \right].
\ee
From here we can determine the {\em degree of coherence} of the initial
superposition, which is defined as the ratio $\Tr\Rcoh / \Tr\rho$
between the trace of the coherent part of the density matrix $\Rcoh$
and the trace of the total density matrix
$\rho = \Rinc + \Rcoh$ ($\Tr\rho \equiv 1$),
\be
\Tr\Rcoh = \sqrt{(\rho_{11}-\rho_{22})^2 + 4\left|\rho_{12}\right|^2}.
\ee
The degree of coherence is equal to the length of the Bloch vector
\cite{Shore90}.
The parameters characterizing the coherent part of the superposition
-- the mixing angle $\mixangle$ and the relative phase $\phase$ --
can be found from the relations
\bml
\label{parameters}
\bea
\tan^2\mixangle &=& \frac{\Rcoh_{22}}{\Rcoh_{11}}, \\
\phase &=& -\arg \Rcoh_{12}.
\eea
\eml
It is easy to verify that in the case of a pure state ($\Ri=0$),
Eqs.~(\ref{parameters}) reduce to Eqs.~(\ref{parameters0}).


\section{Conclusion}

\label{Sec-conclusion}

In this paper, we have proposed a method for measuring the parameters of
a coherent superposition of two atomic states.
It is based upon coupling the two states to a third excited state
by means of two laser pulses, thus mapping the superposition parameters
onto the population of this state.
By measuring the total fluorescence signal from this state for
{\em four} different polarizations of the two pulses we can determine
unambiguously the parameters of the initial superposition --
the populations, the relative phase between the participating states,
and the degree of coherence.
The method requires a proper control of the polarizations and the
relative phase between the two laser fields.
In addition, the two laser pulses have to be short compared to the
lifetime of the excited state in order to avoid spontaneous emission
back to the superposition that would introduce additional incoherence.
On the other hand, the method has the advantage that no knowledge is
required of the exact probability of transition to the excited state;
this makes it invariant against laser power, detuning, and pulse
shape, as long as these are kept constant during the four measurements.
Finally, the method is particularly suited for the frequently used
superpositions between the $m=-1$ and $m=1$ magnetic sublevels
of a $J=1$ level because then only one laser is needed;
this laser can be the same as the one used in the preparation of the
superposition.

\subsection*{Acknowledgements}

BWS thanks the Alexander von Humboldt Stiftung for a Research Award;
his work is supported in part under the auspices
of the U.S. Department of Energy at Lawrence Livermore National
Laboratory under contract W-7405-Eng-48.
RGU thanks the Alexander von Humboldt Stiftung for a Fellowship.



\end{document}